\documentclass[%
 reprint,
 amsmath,amssymb,
 aps,
]{revtex4-2}

\usepackage{graphicx}
\usepackage{dcolumn}
\usepackage{bm}
\usepackage{upgreek}
\DeclareMathOperator*{\argmin}{arg\,min}
\newcommand{\norm}[1]{\left\lVert#1\right\rVert}

\usepackage{hyperref}

\begin{document}

\title[]{Predicting nonlinear optical scattering with physics-driven neural networks}
\author{C. Gigli}
 \email{carlo.gigli@epfl.ch.}
\author{A. Saba}%
\author{A.B. Ayoub}%
\author{D. Psaltis}%
\affiliation{ 
Optics Laboratory, École polytechnique fédérale de Lausanne, CH-1015, Lausanne, Switzerland
}%

\date{\today}

\begin{abstract}
Deep neural networks trained on physical losses are emerging as promising surrogates of nonlinear numerical solvers. These tools can predict solutions of Maxwell's equations and compute gradients of output fields with respect to the material and geometrical properties in millisecond times which makes them attractive for inverse design or inverse scattering applications. Here we develop a tunable version of MaxwellNet, a physics driven neural network able to compute light scattering from inhomogenous media with a size comparable with the incident wavelength in the presence of the optical Kerr effect. The weights of the network are dynamically adjusted to take into account the intensity-dependent refractive index of the material.
\end{abstract}

\maketitle

\section{Introduction}

During the last three decades, the evolution of numerical methods in computational electromagnetics underpinned the solution of many common direct and inverse problems in optics with no analytical counterparts. Multiple scattering, sub-wavelength light confinement, or wave-propagation in complex media are examples where computational techniques have been applied. Nowadays, mesh based techniques, such as finite difference time or frequency domain (FDTD/FDFD) and finite elements method (FEM), represent the most widely adopted tools for the design of many photonic devices at the micro and nanoscale. The goal of the optimization task is to determine geometrical and material properties providing a desired objective field, and it entails the resolution of two consecutive problems: a forward step which relates a given structure with its electromagnetic response, and an inverse one which returns the optimal layout generating the target field. In this framework, the design of nonlinear optical devices drew the attention of the scientific community for the realization of ultra-fast switches as well as all-optical transistors and memories\cite{Soljacic2002,Yanik2003,Nozaki2010}, which constitute the building blocks for the development of fast and low-consumption photonic circuitry. Initially, topology optimization was primarily addressed with evolutionary and gradient-based approaches\cite{Molesky2018,Genty2021}, both leveraging on the numerical solution of several electromagnetic problems for different parameters through the above mentioned tools. In particular adjoint methods \cite{Elesin2012,Lalau-Keraly2013,Hughes2018a} have been shown as a promising solution to compute the derivative of an objective function through the solution of two numerical problems per iteration. 
\par A different emerging approach consists of mapping the relationship between refractive index and electromagnetic field with a data-driven deep neural network (DNN) acting as a surrogate model of numerical methods \cite{Jiang2021a, Ma2021a}. In most cases DNNs act as interpolators which enable us to rapidly predict electromagnetic fields and compute the derivatives of objective functions through back-propagation. The optimization algorithms rely on the same automatic differentiation libraries used in adjoint methods \cite{Minkov2020}, but the inference time of DNNs is typically much faster than nonlinear equation solvers in numerical models. Most of the initial works resorted to a data-driven training through a dataset of input-output pairs computed numerically \cite{Peurifoy2018,Liu2018c,Nadell2019,An2020}. However, this approach ends up being doubly inconvenient: on one side it requires a large amount of simulations for the dataset preparation, on the other hand the knowledge of Maxwell's equations is completely lost behind the data-driven training. To mitigate this effect, in Ref.~\onlinecite{Chen2021Fan} the authors use a hybrid data- and physics-augmented training introducing a regularizer based on Maxwell's equation. 
\par Recently physics-informed neural networks \cite{Lu2021} have been proposed for the solution of Maxwell's equations for inverse scattering problems \cite{Chen2020c,Chen2021a}. This works took inspiration from a seminal paper from Lagaris et al. \cite{Lagaris1998} who highlighted the similarity between neural network training and solving partial differential equations. In this case the network maps spatial and time coordinates to the electromagnetic field in these points. Computing the error on the network output involves the calculation of the gradients with respect to the inputs. Auto-differentiation algorithms are used therefore to rapidly compute the gradients of electric fields with respect to spatial coordinates and consequently solve Maxwell's equations by setting material properties, e.g. permittivity and permeability distributions, and proper boundary conditions. It is worthy to note, however, that this method returns the solution of an individual problem and not a class of problems. 
\par In this work we use MaxwellNet\cite{Lim2021}, a physics driven neural network we recently reported, based on the minimization of Maxwell's equation residual on a class of refractive index distributions. Differently from Refs.\onlinecite{Chen2020c,Chen2021a}, in this case the input of the network is the refractive index distribution $n_0$, and the gradients of electric fields to construct the physical loss are computed through finite difference method. We demonstrate a tunable DNN which can predict very rapidly (in milliseconds time with GPU acceleration) the light scattering from micron-sized features in the presence of the nonlinear AC-Kerr effect for variable incident powers. The method is benchmarked on a set of 2D refractive index distributions for both TE and TM incident polarizations. The results are compared with FEM simulations, exhibiting a few-percentage relative error. It is important to underline that computing gradients through adjoint method or backpropagation algorithms is mathematically equivalent, the objective of the paper is therefore to investigate the possibility to speed up differentiable forward models which take into account optical nonlinearities.

\section{Tunable MaxwellNet: a DNN to solve nonlinear Maxwell equations}

The working principle of MaxwellNet\cite{Lim2021} can be briefly summarized as follows: we feed a DNN based on a U-Net architecture \cite{Ronneberger} with a dataset of 2D refractive index distributions $n(z,x)$ and we get as an output the complex scattered fields for a given incident beam. During the training process, we iteratively update the weights of the U-Net, such that the predicted output fields from the network satisfy Maxwell's equations, instead of comparing the output with ground-truth results from a numerical solver as is generally done in a data-driven process. The advantage of this indirect training using the residual of Maxwell's equations as a physical metric is two-fold: (1) it doesn't require a large amount of numerical simulations to create a database of input-output pairs for the training, consequently reducing the computational costs, and (2) the network directly learns the physics of the system offering a better efficiency in terms of generalization. It should be also noted that in our implementation of physics-informed neural networks, the calculation of the derivatives of the field with respect to the independent variables ($x$, $z$) is done using finite-difference scheme. In contrast, in other schemes the neural network is fed with the independent variables to calculate the derivatives using the chain rule \cite{Lu2021}.  

\begin{figure*}
\centering\includegraphics[width=.65\textwidth]{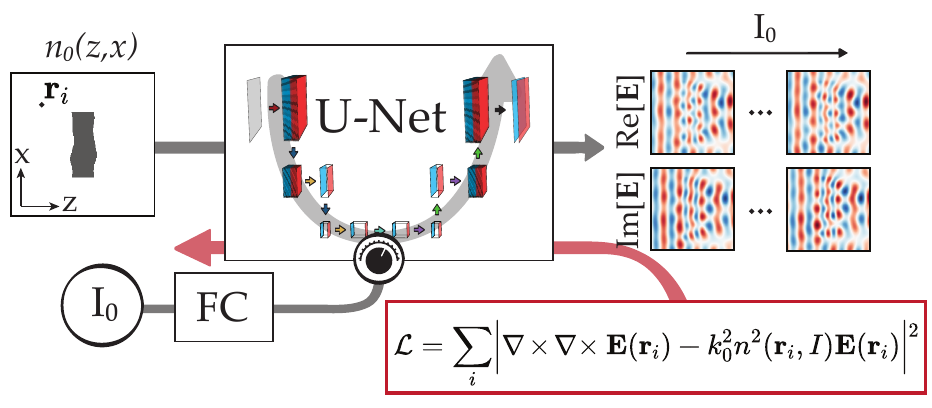}
\caption{Tunable MaxwellNet predicting AC-Kerr scattering. Grey (red) arrows denote forward (backward) propagation. The model is based on a U-Net architecture taking in input the 2D refractive distribution of the material in the $zx$-plane $n(z,x)$ and returning the complex scattered field. By changing the incident beam intensity $I_0$, the weights of specific convolutional layers in the U-Net architecture are tuned so that the physical loss is simultaneously minimized for different working powers.}
\label{fig:sketch}
\end{figure*}

Here we extend MaxwellNet to materials with a non-negligible $\chi^{(3)}$ susceptibility, for which the AC-Kerr effect provide a significant contribution, by using a network that is tunable with the intensity of light. The concept of the work is sketched in Fig.~\ref{fig:sketch}. We start from a set of 2D linear refractive index distributions $n_0(z,x)$, which are discretized on a uniform rectangular grid of size $N_z\times N_x$, and we want to retrieve the scattered field for an incident plane wave propagating along the $z$-axis $\mathbf{E}_i(z,x)=E_0\exp[jk_0z]\hat{\mathbf{u}}$, where $E_0$ is the field amplitude and $I_0=E_0^2/2\eta_0$ its intensity, $k_0$ the wavevector in the background material, and $\hat{\mathbf{u}}$ the polarization unit vector being either $\hat{\mathbf{x}}$ (TM) or $\hat{\mathbf{y}}$ (TE) in the following. We also assume that the value of $\chi^{(3)}$ follows the same spatial distribution as $n_0(z,x)$. 
\par When looking for a function to map a real $N_z\times N_x$ array to a complex one with the same dimensions, the choice of the U-Net architecture is suitable as its structure returns images with the same size as the input while extracting the main features and encode them in a lower dimensional space. In detail, the U-Net block in Fig.~\ref{fig:sketch} encodes the refractive index distributions through 6 convolutional and pooling layers to a latent space of size $N_z/2^5\times N_x/2^5\times512$ and successively decodes them to an output of size $N_z\times N_x\times N_c$ where the $N_c$ channels represent the real and imaginary parts of the scattered field $\mathbf{E}_S(z,x)$. Then, the total field is computed as the sum of the network output and the background field, $\mathbf{E}=\mathbf{E}_S+\mathbf{E}_i$. Weight normalization and leaky ReLU activation functions are adopted at each step. Please refer to the Supplementary Material for further description of the network.
In absence of birefringence, we can consider that the only non-zero components of $\mathbf{E}$ are $E_y$ ($N_c=2$) for the TE and $E_x$, $E_z$ ($N_c=4$) for the TM case, respectively \cite{Snyder1983}. 

The physics-driven training of the network relies on the minimization of Maxwell's equation residual 

\begin{eqnarray}
    \mathcal{L}(n_0,I_0,\pmb{\theta}) = \sum_i& \big\lVert \nabla\times\nabla\times\mathbf{E}(\mathbf{r}_i,I_0,\pmb{\theta})\nonumber\\
    & -k_0^2n^2(\mathbf{r}_i,I)\mathbf{E}(\mathbf{r}_i,I_0,\pmb{\theta})\big\rVert^2 \label{eqn:physloss}
\end{eqnarray}

where the summation is done over all the discretized coordinates in the simulation domain $\mathbf{r}_i$ and $\pmb{\theta}$ includes the weights and the biases of the network. The dependence of $\mathcal{L}$ on $I_0$ in the presence of the AC-Kerr effect is explicitly indicated as the refractive index of the material $n(z,x)$ in \eqref{eqn:physloss} is modulated by the local field intensity $I(r)$:

\begin{equation}
    n = n_0 + \frac{3\chi^{(3)}}{8n_0}|\mathbf{E}(r)|^2 = n_0+n_2 I(r)
    \label{eqn:Kerr_index}
\end{equation}

The numerical discretization of the loss function represents a central point of this work to efficiently train the network. The differential operators in \eqref{eqn:physloss} are calculated with finite differences scheme, and the electromagnetic fields are evaluated on the Yee grid \cite{Yee1966}, the most established method for finite difference time and frequency simulations. Also, the simulation domain is surrounded by a perfectly matching layer (PML) which absorbs the outgoing waves and ensures the fulfillment of Sommerfeld radiation condition \cite{Chew1994,Johnson2021}. Finally, even though computing electric and magnetic fields on the staggered Yee grid improves convergence, discretizing refractive index distributions with sharp transitions on a uniform grid can lead to difficulties in modeling small features and degrade numerical results \cite{Farjadpour2006}. We use a smoothing scheme to alleviate numerical inaccuracies around discontinuous interfaces. The choice of finite difference scheme is very convenient in this context as, more than being a very established method in computational electrodynamics, it can also be easily implemented through 2D convolutions and in turn very well suited for the integration with available machine learning packages as TensorFlow and PyTorch. Please refer to Supplementary Material for further details on the numerical evaluation of \eqref{eqn:physloss}.

We can train the network on a defined set of refractive index distributions for a given incident power, but, as soon as we modify the intensity, the network will no longer be accurate as the physical loss depends on the intensity through $n$. In order address the issue we took inspiration from tunable U-Net architectures \cite{Kang2021} and introduced a fully connected (FC) network (see Fig.~\ref{fig:sketch}) which takes as an input the intensity of the incident plane wave $I_0$ and whose output controls some of the convolutional kernels of the main U-Net. Specifically, a 2D convolutional layer with kernel size $k_{s}\times k_{s}$ going from a tensor with $C_i$ channels to another with $C_o$ channels requires $C_i\times C_o \times k_{s}\times k_{s}$ weights and $C_o$ biases. We therefore set the output dimension of the FC module equal to $C_i\times C_o \times k_{s}\times k_{s} + C_o$ to dynamically tune the second convolutional block in the encoding branch of the U-Net. The choice to control this specific convolutional layer proved to be a good compromise to ensure a good expressivity of the network with a reasonable increase in the number of parameters. The network was implemented in PyTorch 1.7.1 and has a total of $\sim 8$ million parameters. 

\section{Results and discussion}
We tested the ability of MaxwellNet to predict nonlinear scattering in 2D air-suspended glass diffusers with irregular interfaces (Fig.~\ref{fig:result_te}a) excited with a plane wave at $\lambda=1030$ nm propagating along $z$ axis. As a material we consider a standard optical borosilicate glass with linear and second order refractive indices equal to $n_0=1.517$ and $n_2=3.59\times 10^{-20} \textrm{ m}^2/\textrm{W}$, respectively \cite{Weber2003,Adair1987}. The diffusers are 4 $\upmu$m in width and have an average thickness of $2\lambda/n_0$. For such dimensions and choice of parameters, many semi-analytical models, such as beam propagation method, fail to predict accurate solutions due to strong reflections and interference effects inside the material. In these cases one has to resort to fully vectorial simulations, based for example on finite elements or finite differences, which are more demanding from a computational point of view and not well suited for solving inverse design or inverse scattering problems. Furthermore, in presence of self modulation, as in this case, iterative algorithms have to be implemented, which consist in updating the refractive index and the field at each step until convergence, consequently accentuating the burden of numerical computation. 

\par Here we created a dataset $\mathcal{B}$ composed of 5000 $n_0$ distributions (4000, $\mathcal{B}_t$, used for training and 1000, $\mathcal{B}_v$, for validation) and associated to each of them a random input intensity $I_0$. The diffusers' interfaces are generated with low-pass filtered random distributions and the resulting $n_0(z,x)$ functions are discretized on a grid with $N_z=N_x=256$ corresponding to a pixel size of $\lambda/30$. Although all the diffusers have the same average thickness, this is a particularly interesting dataset to benchmark MaxwellNet as the randomness of the interfaces lead to very different scattering properties. 

\subsection{TE case}
As a first example we consider TE polarized incident fields $\mathbf{E}_i(z,x)=E_0\exp[jk_0z]\hat{\mathbf{y}}$ with intensities $I_0$ in the range $[2,10]\times10^{17} \textrm{ W}/\textrm{m}^2$. The lower bound is fixed at the value where the Kerr effect is no longer negligible ($n_2I_0\approx0.005n_0$) and the upper bound corresponds to the maximum value for which we could find a convergent solution with a commercially available numerical solver (COMSOL Multiphysics).

\begin{figure*}
\centering\includegraphics[width=.9\textwidth]{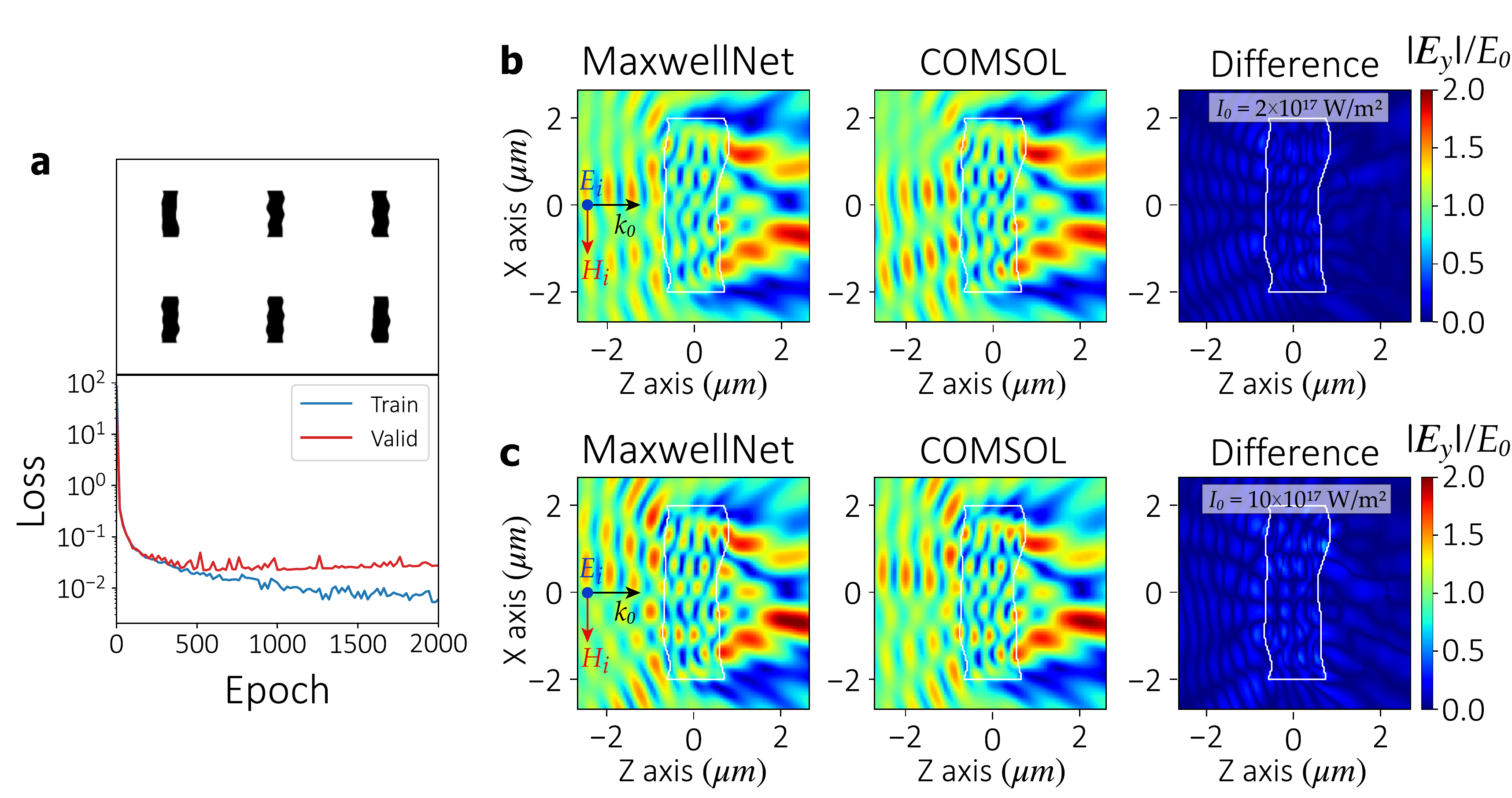}
\caption{MaxwellNet predictions of scattered field from glass diffusers in presence of AC-Kerr effect. (a) Physical loss evaluation for the train (blue) and validation (red) sets. Inset: six examples from the train set. (b) Comparison between the absolute value of total field y-component $E_y$ predicted by MaxwellNet (left), COMSOL (center) and their difference (right) for an incident intensity $I_0=2\times10^{17} \textrm{ W}/\textrm{m}^2$. (c) Same as in (b) but for an incident intensity $I_0=10\times10^{17} \textrm{ W}/\textrm{m}^2$. The same element in the validation set is considered in (b) and (c). The electric field values are normalized by the incident amplitude $E_0$.}
\label{fig:result_te}
\end{figure*}

The training was performed using mini batches of size 8 and the Adam optimizer with a scheduled learning rate starting from an initial value $l_r=2\times10^{-4}$ and decreasing of 30\% every 1000 epochs.
During the training the optimizer looks for the set of parameters $\hat{\pmb{\theta}}$ satisfying:
\begin{equation}
    \hat{\pmb{\theta}} = \argmin_{\pmb{\theta}} \sum_{(n_0,I_0)\in\mathcal{B}_t} 
    \mathcal{L}(n_0,I_0,\pmb{\theta})
    \label{eqn:train}
\end{equation}

\begin{figure}[h!]
\centering\includegraphics[width=.5\textwidth]{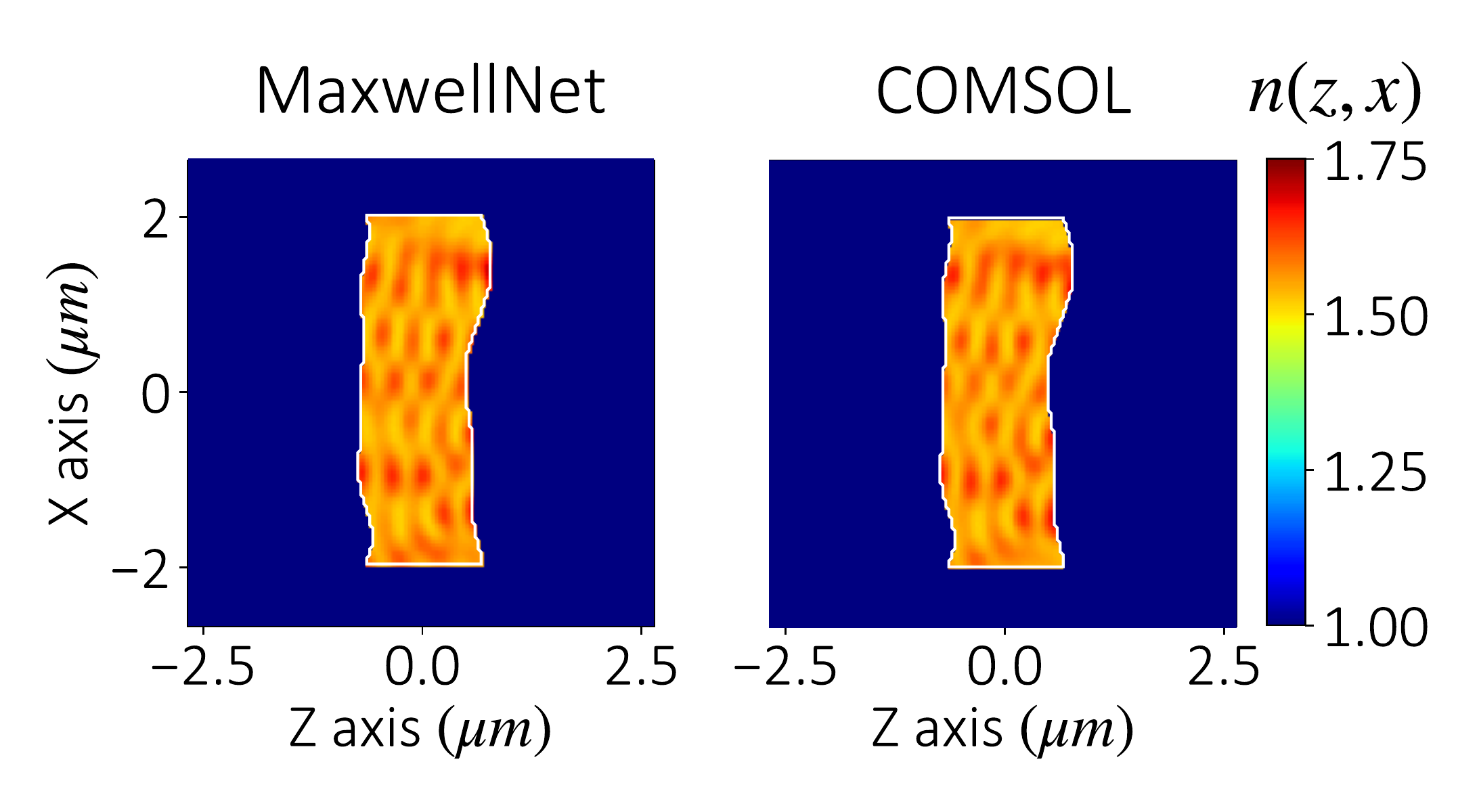}
\caption{Nonlinear refractive index $n(z,x)$ computed as in \eqref{eqn:Kerr_index} predicted by MaxwellNet (left) and calculated with COMSOL (right) for the same sample as in Fig.\ref{fig:result_te}.} 
\label{fig:index_te}
\end{figure}

Fig.~\ref{fig:result_te}a shows that the loss computed on the validation set $\mathcal{B}_v$ starts saturating after 800 epochs, corresponding to a training time of about 8 hours on a machine with GPU acceleration (Nvidia Tesla V100-32GB). Fig.~\ref{fig:result_te}b and c compare the absolute value of the field predictions from MaxwellNet and COMSOL Multiphysics for the same element in the validation set at incident intensities $I_0=2\times10^{17} \textrm{ W}/\textrm{m}^2$ (b) and $I_0=10\times10^{17} \textrm{ W}/\textrm{m}^2$ (c), respectively. Looking at the field inside the diffusers and the reflections we can confirm that MaxwellNet accurately captures the intensity dependent behavior of the scattered field. We do attribute the remaining discrepancies between MaxwellNet and COMSOL to discretization and the limited capacity of the network. Fig.~\ref{fig:index_te} reports also the comparison between the modulated refractive index computed as in \eqref{eqn:Kerr_index} with MaxwellNet and COMSOL for an incident intensity $I_0=10\times10^{17} \textrm{ W}/\textrm{m}^2$. We can notice in these conditions a refractive index modulation of about $0.17\times n_0$ occurring on a spatial distance of $0.1\times\lambda/n_0$ which reflects COMSOL predictions within a maximum discrepancy of 2\%. A further comparison between the refractive index predicted by MaxwellNet and COMSOL considering or neglecting Kerr effect is provided in the Supplementary Material. 

\begin{figure*}
\centering\includegraphics[width=.98\textwidth]{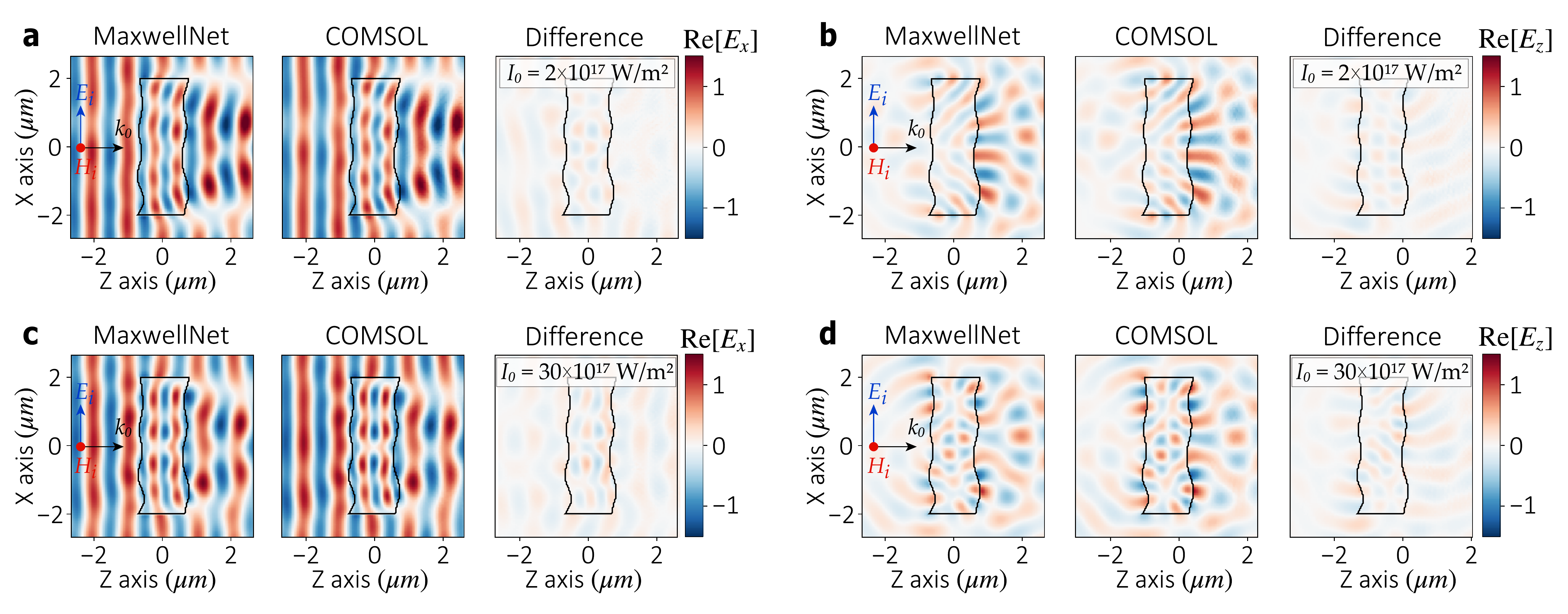}
\caption{MaxwellNet predictions of scattered field from a glass diffuser in $\mathcal{B}_v$ in presence of AC-Kerr effect for TM polarization. (a-b) Real part of the total field x-(a) and z-component(b) predicted by MaxwellNet (left) or computed with COMSOL (center) and their difference (right). The blue, red and black arrows denote the electric, magnetic field and wavevector of the incident beam. The results refer to an incident power $I_0=2\times10^{17}\textrm{ W}/\textrm{m}^2$. (c-d) Same as in (a-b) but for an incident power $I_0=30\times10^{17}\textrm{ W}/\textrm{m}^2$. Fields are normalized by the incident amplitude $E_0$ in each case.}
\label{fig:result_tm}
\end{figure*}

\subsection{TM case}
The electric field in \eqref{eqn:physloss} was written as a vector, but for TE case in 2D, Maxwell's equation residual can be rewritten in a scalar form as the only non-zero component of the electric field is $E_y$. In order to extend the proposed method a fully vectorial case we consider as a second example the same dataset of diffusers excited by a TM polarized plane wave $\mathbf{E}_i(z,x)=E_0\exp[jk_0z]\hat{\mathbf{x}}$ with intensities in the range $[2,30]\times10^{17} \textrm{ W}/\textrm{m}^2$. The network has the same structure as the previous case except for the last convolution which now returns a tensor with $N_c=4$ channels being the real and imaginary parts of $E_x$ and $E_z$ which are the only non-zero components of the total electric field. The validation loss saturated in this case after 2000 epochs, corresponding to a training time of about 20 hours on the same machine as before. 

Fig.~\ref{fig:result_tm} reports the comparison of MaxwellNet predictions and ground truth computed in COMSOL for a sample in the validation set $\mathcal{B}_v$ considering the two limiting values of the intensity range used during training. Instead of the amplitude, we show in this case the real part of $E_x$ and $E_z$ which enables to highlight the faster oscillations inside the glass at high power (a-c) and the strong modulation of the cross-polarized field (b-d). In both cases MaxwellNet predictions are consistent with finite element simulations. For a comparison with prediction neglecting Kerr effect and a video of MaxwellNet inference increasing incident power please refer to Supplementary Material.

In order to evaluate the suitability of MaxwellNet as a forward model to solve scattering problems or as a tool for inverse design we should point out few considerations in terms of speed and accuracy. Although the long training time required by MaxwellNet might seem excessive for any useful application, on the other hand the inference time is much shorter compared to commercially available solvers. Concerning the accuracy, we quantitatively evaluate the performance of MaxwellNet by defining the relative error with respect to the ground truth solution as:

\begin{equation}
    \varepsilon_r=\frac{\iint\norm{\mathbf{E}^M(z,x)-\mathbf{E}^C(z,x)}^2dzdx}{\iint\norm{\mathbf{E}^C(z,x)}^2dzdx}
    \label{eqn:error}
\end{equation}

where $\mathbf{E}^M$ and $\mathbf{E}^C$ are the total field computed with MaxwellNet and COMSOL, respectively, and the integration is restricted to the physical domain excluding the PML regions. Table \ref{tab:performance} summarizes speed and accuracy performance of MaxwellNet compared to COMSOL. At large powers, for which COMSOL iterative scheme takes longer times to converge, MaxwellNet provides an acceleration factor of $\sim10^3$ on a common workstation with 3.40 GHz CPU and 8 GB GPU (NVIDIA GeForce GTX 1070) with a maximum relative error of about 4\%.

\begin{table}[]
    \centering
    \begin{tabular}{l c c}
    \hline\hline
         & TE &  TM\\
         \hline
         COMSOL solution time & 5-17s & 5-20s\\
         MaxwellNet training & 8h& 20h\\
         MaxwellNet inference & 15ms & 21ms\\
         MaxwellNet error $\varepsilon_r$ & $1.7-3.0 \times 10^{-2}$& $2.5-3.9 \times 10^{-2}$\\
    \hline\hline
    \end{tabular}
    \caption{Evaluation of MaxwellNet performances compared to COMSOL. The reported intervals refer to the lowest and highest powers used in the dataset.}
    \label{tab:performance}
\end{table}

\section{Conclusion}

We described a physically-driven training of a DNN which can predict nonlinear optical scattering from micro-objects with a size of few wavelengths. There are at least three noteworthy aspects of the present implementation. The first one is the integration of a physical loss including nonlinear optical effects within PyTorch framework. The calculation of this physical metric through finite difference scheme on the Yee grid results in numerically accurate and very convenient calculations through 2D convolution tools available in most of machine learning packages. Furthermore, the same criterion might also be applied as physical regularizer in many minimization tasks or as a prior in inverse scattering problems. 
\par Most significantly, we proposed a forward model which can infer the scattered field 1000 times faster than numerical solver for the problem we considered and whose output can be rapidly back-propagated through auto-differentiation techniques. 
\par Finally, we demonstrated a tunable network which can solve nonlinear scattering problem for different working conditions. On the one hand this is a necessary tool for applications as the inverse design of nonlinear optical devices and on the other it points to the possibility of introducing more control parameters, such as working wavelength, incident angle and so on, to achieve broadband designs. Such a network is well adapted for the integration with generative models for the reparametrization of photonic devices design \cite{Jiang2019c,Lim2021,Chen2021b}. 
\par We can identify some critical points in the current implementation. First, the numerical accuracy strongly depends on the discretization size. We expect that for more complex structures or sharp high-index transitions more refined finite differences techniques should be implemented to improve the numerical accuracy. Second, it is important to point out that there is a key difference between the local intensity, $I$, which modulates the refractive index in \eqref{eqn:Kerr_index}, and the incident one, $I_0$. In particular, for strongly resonant systems these two can differ significantly and tuning the network only through $I_0$ might not be enough to capture the physics of the system. Furthermore, as it happens for most neural networks, the model offers promising results in interpolation, i.e. when inferring on refractive index distributions similar to the training dataset, but the performances drop when feeding the network with inputs statistically different from those seen during training. 
\par However we acknowledge these as current technical issues and they do not constitute fundamental limitations of the method. We do therefore expect that the advantages of indirect physical-driven training combined with increasing computational power of GPUs will enable in the coming years to the extension of this method to full vectorial 3D models with fine spatial resolution which will constitute useful tools for the simulation and inverse design of nonlinear optical devices as ultra-fast switches, nonlinear photonic cavities or nonlinear metasurfaces.

\section*{Disclosures}
The authors declare no competing interests.

\begin{acknowledgments}
The authors thank Dr. Joowon Lim for fruitful discussions. This project was funded by the Swiss National Science Foundation (SNSF) funding number 514481. 
\end{acknowledgments}


\bibliography{biblio}

\begin{thebibliography}{32}%
\makeatletter
\providecommand \@ifxundefined [1]{%
 \@ifx{#1\undefined}
}%
\providecommand \@ifnum [1]{%
 \ifnum #1\expandafter \@firstoftwo
 \else \expandafter \@secondoftwo
 \fi
}%
\providecommand \@ifx [1]{%
 \ifx #1\expandafter \@firstoftwo
 \else \expandafter \@secondoftwo
 \fi
}%
\providecommand \natexlab [1]{#1}%
\providecommand \enquote  [1]{``#1''}%
\providecommand \bibnamefont  [1]{#1}%
\providecommand \bibfnamefont [1]{#1}%
\providecommand \citenamefont [1]{#1}%
\providecommand \href@noop [0]{\@secondoftwo}%
\providecommand \href [0]{\begingroup \@sanitize@url \@href}%
\providecommand \@href[1]{\@@startlink{#1}\@@href}%
\providecommand \@@href[1]{\endgroup#1\@@endlink}%
\providecommand \@sanitize@url [0]{\catcode `\\12\catcode `\$12\catcode
  `\&12\catcode `\#12\catcode `\^12\catcode `\_12\catcode `\%12\relax}%
\providecommand \@@startlink[1]{}%
\providecommand \@@endlink[0]{}%
\providecommand \url  [0]{\begingroup\@sanitize@url \@url }%
\providecommand \@url [1]{\endgroup\@href {#1}{\urlprefix }}%
\providecommand \urlprefix  [0]{URL }%
\providecommand \Eprint [0]{\href }%
\providecommand \doibase [0]{https://doi.org/}%
\providecommand \selectlanguage [0]{\@gobble}%
\providecommand \bibinfo  [0]{\@secondoftwo}%
\providecommand \bibfield  [0]{\@secondoftwo}%
\providecommand \translation [1]{[#1]}%
\providecommand \BibitemOpen [0]{}%
\providecommand \bibitemStop [0]{}%
\providecommand \bibitemNoStop [0]{.\EOS\space}%
\providecommand \EOS [0]{\spacefactor3000\relax}%
\providecommand \BibitemShut  [1]{\csname bibitem#1\endcsname}%
\let\auto@bib@innerbib\@empty
\bibitem [{\citenamefont {Solja{\v{c}}i{\'{c}}}\ \emph
  {et~al.}(2002)\citenamefont {Solja{\v{c}}i{\'{c}}}, \citenamefont {Ibanescu},
  \citenamefont {Johnson}, \citenamefont {Fink},\ and\ \citenamefont
  {Joannopoulos}}]{Soljacic2002}%
  \BibitemOpen
  \bibfield  {author} {\bibinfo {author} {\bibfnamefont {M.}~\bibnamefont
  {Solja{\v{c}}i{\'{c}}}}, \bibinfo {author} {\bibfnamefont {M.}~\bibnamefont
  {Ibanescu}}, \bibinfo {author} {\bibfnamefont {S.~G.}\ \bibnamefont
  {Johnson}}, \bibinfo {author} {\bibfnamefont {Y.}~\bibnamefont {Fink}},\ and\
  \bibinfo {author} {\bibfnamefont {J.~D.}\ \bibnamefont {Joannopoulos}},\
  }\bibfield  {title} {\bibinfo {title} {{Optimal bistable switching in
  nonlinear photonic crystals}},\ }\href
  {https://doi.org/10.1103/PhysRevE.66.055601} {\bibfield  {journal} {\bibinfo
  {journal} {Physical Review E}\ }\textbf {\bibinfo {volume} {66}},\ \bibinfo
  {pages} {1} (\bibinfo {year} {2002})}\BibitemShut {NoStop}%
\bibitem [{\citenamefont {Yanik}\ \emph {et~al.}(2003)\citenamefont {Yanik},
  \citenamefont {Fan}, \citenamefont {Solja{\v{c}}i{\'{c}}},\ and\
  \citenamefont {Joannopoulos}}]{Yanik2003}%
  \BibitemOpen
  \bibfield  {author} {\bibinfo {author} {\bibfnamefont {M.~F.}\ \bibnamefont
  {Yanik}}, \bibinfo {author} {\bibfnamefont {S.}~\bibnamefont {Fan}}, \bibinfo
  {author} {\bibfnamefont {M.}~\bibnamefont {Solja{\v{c}}i{\'{c}}}},\ and\
  \bibinfo {author} {\bibfnamefont {J.~D.}\ \bibnamefont {Joannopoulos}},\
  }\bibfield  {title} {\bibinfo {title} {{All-optical transistor action with
  bistable switching in a photonic crystal cross-waveguide geometry}},\ }\href
  {https://doi.org/10.1364/ol.28.002506} {\bibfield  {journal} {\bibinfo
  {journal} {Optics Letters}\ }\textbf {\bibinfo {volume} {28}},\ \bibinfo
  {pages} {2506} (\bibinfo {year} {2003})}\BibitemShut {NoStop}%
\bibitem [{\citenamefont {Nozaki}\ \emph {et~al.}(2010)\citenamefont {Nozaki},
  \citenamefont {Tanabe}, \citenamefont {Shinya}, \citenamefont {Matsuo},
  \citenamefont {Sato}, \citenamefont {Taniyama},\ and\ \citenamefont
  {Notomi}}]{Nozaki2010}%
  \BibitemOpen
  \bibfield  {author} {\bibinfo {author} {\bibfnamefont {K.}~\bibnamefont
  {Nozaki}}, \bibinfo {author} {\bibfnamefont {T.}~\bibnamefont {Tanabe}},
  \bibinfo {author} {\bibfnamefont {A.}~\bibnamefont {Shinya}}, \bibinfo
  {author} {\bibfnamefont {S.}~\bibnamefont {Matsuo}}, \bibinfo {author}
  {\bibfnamefont {T.}~\bibnamefont {Sato}}, \bibinfo {author} {\bibfnamefont
  {H.}~\bibnamefont {Taniyama}},\ and\ \bibinfo {author} {\bibfnamefont
  {M.}~\bibnamefont {Notomi}},\ }\bibfield  {title} {\bibinfo {title}
  {{Sub-femtojoule all-optical switching using a photonic-crystal
  nanocavity}},\ }\href {https://doi.org/10.1038/nphoton.2010.89} {\bibfield
  {journal} {\bibinfo  {journal} {Nature Photonics}\ }\textbf {\bibinfo
  {volume} {4}},\ \bibinfo {pages} {477} (\bibinfo {year} {2010})}\BibitemShut
  {NoStop}%
\bibitem [{\citenamefont {Molesky}\ \emph {et~al.}(2018)\citenamefont
  {Molesky}, \citenamefont {Lin}, \citenamefont {Piggott}, \citenamefont {Jin},
  \citenamefont {Vuckovi{\'{c}}},\ and\ \citenamefont
  {Rodriguez}}]{Molesky2018}%
  \BibitemOpen
  \bibfield  {author} {\bibinfo {author} {\bibfnamefont {S.}~\bibnamefont
  {Molesky}}, \bibinfo {author} {\bibfnamefont {Z.}~\bibnamefont {Lin}},
  \bibinfo {author} {\bibfnamefont {A.~Y.}\ \bibnamefont {Piggott}}, \bibinfo
  {author} {\bibfnamefont {W.}~\bibnamefont {Jin}}, \bibinfo {author}
  {\bibfnamefont {J.}~\bibnamefont {Vuckovi{\'{c}}}},\ and\ \bibinfo {author}
  {\bibfnamefont {A.~W.}\ \bibnamefont {Rodriguez}},\ }\bibfield  {title}
  {\bibinfo {title} {{Inverse design in nanophotonics}},\ }\href
  {https://doi.org/10.1038/s41566-018-0246-9} {\bibfield  {journal} {\bibinfo
  {journal} {Nature Photonics}\ }\textbf {\bibinfo {volume} {12}},\ \bibinfo
  {pages} {659} (\bibinfo {year} {2018})}\BibitemShut {NoStop}%
\bibitem [{\citenamefont {Genty}\ \emph {et~al.}(2021)\citenamefont {Genty},
  \citenamefont {Salmela}, \citenamefont {Dudley}, \citenamefont {Brunner},
  \citenamefont {Kokhanovskiy}, \citenamefont {Kobtsev},\ and\ \citenamefont
  {Turitsyn}}]{Genty2021}%
  \BibitemOpen
  \bibfield  {author} {\bibinfo {author} {\bibfnamefont {G.}~\bibnamefont
  {Genty}}, \bibinfo {author} {\bibfnamefont {L.}~\bibnamefont {Salmela}},
  \bibinfo {author} {\bibfnamefont {J.~M.}\ \bibnamefont {Dudley}}, \bibinfo
  {author} {\bibfnamefont {D.}~\bibnamefont {Brunner}}, \bibinfo {author}
  {\bibfnamefont {A.}~\bibnamefont {Kokhanovskiy}}, \bibinfo {author}
  {\bibfnamefont {S.}~\bibnamefont {Kobtsev}},\ and\ \bibinfo {author}
  {\bibfnamefont {S.~K.}\ \bibnamefont {Turitsyn}},\ }\bibfield  {title}
  {\bibinfo {title} {{Machine learning and applications in ultrafast
  photonics}},\ }\href {https://doi.org/10.1038/s41566-020-00716-4} {\bibfield
  {journal} {\bibinfo  {journal} {Nature Photonics}\ }\textbf {\bibinfo
  {volume} {15}},\ \bibinfo {pages} {91} (\bibinfo {year} {2021})}\BibitemShut
  {NoStop}%
\bibitem [{\citenamefont {Elesin}\ \emph {et~al.}(2012)\citenamefont {Elesin},
  \citenamefont {Lazarov}, \citenamefont {Jensen},\ and\ \citenamefont
  {Sigmund}}]{Elesin2012}%
  \BibitemOpen
  \bibfield  {author} {\bibinfo {author} {\bibfnamefont {Y.}~\bibnamefont
  {Elesin}}, \bibinfo {author} {\bibfnamefont {B.~S.}\ \bibnamefont {Lazarov}},
  \bibinfo {author} {\bibfnamefont {J.~S.}\ \bibnamefont {Jensen}},\ and\
  \bibinfo {author} {\bibfnamefont {O.}~\bibnamefont {Sigmund}},\ }\bibfield
  {title} {\bibinfo {title} {{Design of robust and efficient photonic switches
  using topology optimization}},\ }\href
  {https://doi.org/10.1016/j.photonics.2011.10.003} {\bibfield  {journal}
  {\bibinfo  {journal} {Photonics and Nanostructures - Fundamentals and
  Applications}\ }\textbf {\bibinfo {volume} {10}},\ \bibinfo {pages} {153}
  (\bibinfo {year} {2012})}\BibitemShut {NoStop}%
\bibitem [{\citenamefont {Lalau-Keraly}\ \emph {et~al.}(2013)\citenamefont
  {Lalau-Keraly}, \citenamefont {Bhargava}, \citenamefont {Miller},\ and\
  \citenamefont {Yablonovitch}}]{Lalau-Keraly2013}%
  \BibitemOpen
  \bibfield  {author} {\bibinfo {author} {\bibfnamefont {C.~M.}\ \bibnamefont
  {Lalau-Keraly}}, \bibinfo {author} {\bibfnamefont {S.}~\bibnamefont
  {Bhargava}}, \bibinfo {author} {\bibfnamefont {O.~D.}\ \bibnamefont
  {Miller}},\ and\ \bibinfo {author} {\bibfnamefont {E.}~\bibnamefont
  {Yablonovitch}},\ }\bibfield  {title} {\bibinfo {title} {{Adjoint shape
  optimization applied to electromagnetic design}},\ }\href
  {https://doi.org/10.1364/oe.21.021693} {\bibfield  {journal} {\bibinfo
  {journal} {Optics Express}\ }\textbf {\bibinfo {volume} {21}},\ \bibinfo
  {pages} {21693} (\bibinfo {year} {2013})}\BibitemShut {NoStop}%
\bibitem [{\citenamefont {Hughes}\ \emph {et~al.}(2018)\citenamefont {Hughes},
  \citenamefont {Minkov}, \citenamefont {Williamson},\ and\ \citenamefont
  {Fan}}]{Hughes2018a}%
  \BibitemOpen
  \bibfield  {author} {\bibinfo {author} {\bibfnamefont {T.~W.}\ \bibnamefont
  {Hughes}}, \bibinfo {author} {\bibfnamefont {M.}~\bibnamefont {Minkov}},
  \bibinfo {author} {\bibfnamefont {I.~A.}\ \bibnamefont {Williamson}},\ and\
  \bibinfo {author} {\bibfnamefont {S.}~\bibnamefont {Fan}},\ }\bibfield
  {title} {\bibinfo {title} {{Adjoint Method and Inverse Design for Nonlinear
  Nanophotonic Devices}},\ }\href
  {https://doi.org/10.1021/acsphotonics.8b01522} {\bibfield  {journal}
  {\bibinfo  {journal} {ACS Photonics}\ }\textbf {\bibinfo {volume} {5}},\
  \bibinfo {pages} {4781} (\bibinfo {year} {2018})}\BibitemShut {NoStop}%
\bibitem [{\citenamefont {Jiang}\ \emph {et~al.}(2021)\citenamefont {Jiang},
  \citenamefont {Chen},\ and\ \citenamefont {Fan}}]{Jiang2021a}%
  \BibitemOpen
  \bibfield  {author} {\bibinfo {author} {\bibfnamefont {J.}~\bibnamefont
  {Jiang}}, \bibinfo {author} {\bibfnamefont {M.}~\bibnamefont {Chen}},\ and\
  \bibinfo {author} {\bibfnamefont {J.~A.}\ \bibnamefont {Fan}},\ }\bibfield
  {title} {\bibinfo {title} {{Deep neural networks for the evaluation and
  design of photonic devices}},\ }\href
  {https://doi.org/10.1038/s41578-020-00260-1} {\bibfield  {journal} {\bibinfo
  {journal} {Nature Reviews Materials}\ }\textbf {\bibinfo {volume} {6}},\
  \bibinfo {pages} {679} (\bibinfo {year} {2021})}\BibitemShut {NoStop}%
\bibitem [{\citenamefont {Ma}\ \emph {et~al.}(2021)\citenamefont {Ma},
  \citenamefont {Liu}, \citenamefont {Kudyshev}, \citenamefont {Boltasseva},
  \citenamefont {Cai},\ and\ \citenamefont {Liu}}]{Ma2021a}%
  \BibitemOpen
  \bibfield  {author} {\bibinfo {author} {\bibfnamefont {W.}~\bibnamefont
  {Ma}}, \bibinfo {author} {\bibfnamefont {Z.}~\bibnamefont {Liu}}, \bibinfo
  {author} {\bibfnamefont {Z.~A.}\ \bibnamefont {Kudyshev}}, \bibinfo {author}
  {\bibfnamefont {A.}~\bibnamefont {Boltasseva}}, \bibinfo {author}
  {\bibfnamefont {W.}~\bibnamefont {Cai}},\ and\ \bibinfo {author}
  {\bibfnamefont {Y.}~\bibnamefont {Liu}},\ }\bibfield  {title} {\bibinfo
  {title} {{Deep learning for the design of photonic structures}},\ }\href
  {https://doi.org/10.1038/s41566-020-0685-y} {\bibfield  {journal} {\bibinfo
  {journal} {Nature Photonics}\ }\textbf {\bibinfo {volume} {15}},\ \bibinfo
  {pages} {77} (\bibinfo {year} {2021})}\BibitemShut {NoStop}%
\bibitem [{\citenamefont {Minkov}\ \emph {et~al.}(2020)\citenamefont {Minkov},
  \citenamefont {Williamson}, \citenamefont {Andreani}, \citenamefont {Gerace},
  \citenamefont {Lou}, \citenamefont {Song}, \citenamefont {Hughes},\ and\
  \citenamefont {Fan}}]{Minkov2020}%
  \BibitemOpen
  \bibfield  {author} {\bibinfo {author} {\bibfnamefont {M.}~\bibnamefont
  {Minkov}}, \bibinfo {author} {\bibfnamefont {I.~A.}\ \bibnamefont
  {Williamson}}, \bibinfo {author} {\bibfnamefont {L.~C.}\ \bibnamefont
  {Andreani}}, \bibinfo {author} {\bibfnamefont {D.}~\bibnamefont {Gerace}},
  \bibinfo {author} {\bibfnamefont {B.}~\bibnamefont {Lou}}, \bibinfo {author}
  {\bibfnamefont {A.~Y.}\ \bibnamefont {Song}}, \bibinfo {author}
  {\bibfnamefont {T.~W.}\ \bibnamefont {Hughes}},\ and\ \bibinfo {author}
  {\bibfnamefont {S.}~\bibnamefont {Fan}},\ }\bibfield  {title} {\bibinfo
  {title} {{Inverse Design of Photonic Crystals through Automatic
  Differentiation}},\ }\href {https://doi.org/10.1021/acsphotonics.0c00327}
  {\bibfield  {journal} {\bibinfo  {journal} {ACS Photonics}\ }\textbf
  {\bibinfo {volume} {7}},\ \bibinfo {pages} {1729} (\bibinfo {year}
  {2020})}\BibitemShut {NoStop}%
\bibitem [{\citenamefont {Peurifoy}\ \emph {et~al.}(2018)\citenamefont
  {Peurifoy}, \citenamefont {Shen}, \citenamefont {Jing}, \citenamefont {Yang},
  \citenamefont {Cano-Renteria}, \citenamefont {DeLacy}, \citenamefont
  {Joannopoulos}, \citenamefont {Tegmark}, \citenamefont
  {Solja{\v{c}}i{\'{c}}}, \citenamefont {Joannopoulos},\ and\ \citenamefont
  {Soljaci{\'{c}}}}]{Peurifoy2018}%
  \BibitemOpen
  \bibfield  {author} {\bibinfo {author} {\bibfnamefont {J.}~\bibnamefont
  {Peurifoy}}, \bibinfo {author} {\bibfnamefont {Y.}~\bibnamefont {Shen}},
  \bibinfo {author} {\bibfnamefont {L.}~\bibnamefont {Jing}}, \bibinfo {author}
  {\bibfnamefont {Y.}~\bibnamefont {Yang}}, \bibinfo {author} {\bibfnamefont
  {F.}~\bibnamefont {Cano-Renteria}}, \bibinfo {author} {\bibfnamefont {B.~G.}\
  \bibnamefont {DeLacy}}, \bibinfo {author} {\bibfnamefont {J.~D.}\
  \bibnamefont {Joannopoulos}}, \bibinfo {author} {\bibfnamefont
  {M.}~\bibnamefont {Tegmark}}, \bibinfo {author} {\bibfnamefont
  {M.}~\bibnamefont {Solja{\v{c}}i{\'{c}}}}, \bibinfo {author} {\bibfnamefont
  {J.~D.}\ \bibnamefont {Joannopoulos}},\ and\ \bibinfo {author} {\bibfnamefont
  {M.}~\bibnamefont {Soljaci{\'{c}}}},\ }\bibfield  {title} {\bibinfo {title}
  {{Nanophotonic particle simulation and inverse design using artificial neural
  networks}},\ }\href {https://doi.org/10.1126/sciadv.aar4206} {\bibfield
  {journal} {\bibinfo  {journal} {Science Advances}\ }\textbf {\bibinfo
  {volume} {4:eaar4206}},\ \bibinfo {pages} {1} (\bibinfo {year}
  {2018})}\BibitemShut {NoStop}%
\bibitem [{\citenamefont {Liu}\ \emph {et~al.}(2018)\citenamefont {Liu},
  \citenamefont {Zhu}, \citenamefont {Rodrigues}, \citenamefont {Lee},\ and\
  \citenamefont {Cai}}]{Liu2018c}%
  \BibitemOpen
  \bibfield  {author} {\bibinfo {author} {\bibfnamefont {Z.}~\bibnamefont
  {Liu}}, \bibinfo {author} {\bibfnamefont {D.}~\bibnamefont {Zhu}}, \bibinfo
  {author} {\bibfnamefont {S.~P.}\ \bibnamefont {Rodrigues}}, \bibinfo {author}
  {\bibfnamefont {K.~T.}\ \bibnamefont {Lee}},\ and\ \bibinfo {author}
  {\bibfnamefont {W.}~\bibnamefont {Cai}},\ }\bibfield  {title} {\bibinfo
  {title} {{Generative Model for the Inverse Design of Metasurfaces}},\ }\href
  {https://doi.org/10.1021/acs.nanolett.8b03171} {\bibfield  {journal}
  {\bibinfo  {journal} {Nano Letters}\ }\textbf {\bibinfo {volume} {18}},\
  \bibinfo {pages} {6570} (\bibinfo {year} {2018})}\BibitemShut {NoStop}%
\bibitem [{\citenamefont {Nadell}\ \emph {et~al.}(2019)\citenamefont {Nadell},
  \citenamefont {Huang}, \citenamefont {Malof},\ and\ \citenamefont
  {Padilla}}]{Nadell2019}%
  \BibitemOpen
  \bibfield  {author} {\bibinfo {author} {\bibfnamefont {C.~C.}\ \bibnamefont
  {Nadell}}, \bibinfo {author} {\bibfnamefont {B.}~\bibnamefont {Huang}},
  \bibinfo {author} {\bibfnamefont {J.~M.}\ \bibnamefont {Malof}},\ and\
  \bibinfo {author} {\bibfnamefont {W.~J.}\ \bibnamefont {Padilla}},\
  }\bibfield  {title} {\bibinfo {title} {{Deep learning for accelerated
  all-dielectric metasurface design}},\ }\href
  {https://doi.org/10.1364/oe.27.027523} {\bibfield  {journal} {\bibinfo
  {journal} {Optics Express}\ }\textbf {\bibinfo {volume} {27}},\ \bibinfo
  {pages} {27523} (\bibinfo {year} {2019})}\BibitemShut {NoStop}%
\bibitem [{\citenamefont {An}\ \emph {et~al.}(2020)\citenamefont {An},
  \citenamefont {Zheng}, \citenamefont {Shalaginov}, \citenamefont {Tang},
  \citenamefont {Li}, \citenamefont {Zhou}, \citenamefont {Ding}, \citenamefont
  {Agarwal}, \citenamefont {Rivero-Baleine}, \citenamefont {Kang},
  \citenamefont {Richardson}, \citenamefont {Gu}, \citenamefont {Hu},
  \citenamefont {Fowler},\ and\ \citenamefont {Zhang}}]{An2020}%
  \BibitemOpen
  \bibfield  {author} {\bibinfo {author} {\bibfnamefont {S.}~\bibnamefont
  {An}}, \bibinfo {author} {\bibfnamefont {B.}~\bibnamefont {Zheng}}, \bibinfo
  {author} {\bibfnamefont {M.~Y.}\ \bibnamefont {Shalaginov}}, \bibinfo
  {author} {\bibfnamefont {H.}~\bibnamefont {Tang}}, \bibinfo {author}
  {\bibfnamefont {H.}~\bibnamefont {Li}}, \bibinfo {author} {\bibfnamefont
  {L.}~\bibnamefont {Zhou}}, \bibinfo {author} {\bibfnamefont {J.}~\bibnamefont
  {Ding}}, \bibinfo {author} {\bibfnamefont {A.~M.}\ \bibnamefont {Agarwal}},
  \bibinfo {author} {\bibfnamefont {C.}~\bibnamefont {Rivero-Baleine}},
  \bibinfo {author} {\bibfnamefont {M.}~\bibnamefont {Kang}}, \bibinfo {author}
  {\bibfnamefont {K.~A.}\ \bibnamefont {Richardson}}, \bibinfo {author}
  {\bibfnamefont {T.}~\bibnamefont {Gu}}, \bibinfo {author} {\bibfnamefont
  {J.}~\bibnamefont {Hu}}, \bibinfo {author} {\bibfnamefont {C.}~\bibnamefont
  {Fowler}},\ and\ \bibinfo {author} {\bibfnamefont {H.}~\bibnamefont
  {Zhang}},\ }\bibfield  {title} {\bibinfo {title} {{Deep learning modeling
  approach for metasurfaces with high degrees of freedom}},\ }\href
  {https://doi.org/10.1364/oe.401960} {\bibfield  {journal} {\bibinfo
  {journal} {Optics Express}\ }\textbf {\bibinfo {volume} {28}},\ \bibinfo
  {pages} {31932} (\bibinfo {year} {2020})}\BibitemShut {NoStop}%
\bibitem [{\citenamefont {Chen}\ \emph {et~al.}(2021)\citenamefont {Chen},
  \citenamefont {Lupoiu}, \citenamefont {Mao}, \citenamefont {Huang},
  \citenamefont {Jiang}, \citenamefont {Lalanne},\ and\ \citenamefont
  {Fan}}]{Chen2021Fan}%
  \BibitemOpen
  \bibfield  {author} {\bibinfo {author} {\bibfnamefont {M.}~\bibnamefont
  {Chen}}, \bibinfo {author} {\bibfnamefont {R.}~\bibnamefont {Lupoiu}},
  \bibinfo {author} {\bibfnamefont {C.}~\bibnamefont {Mao}}, \bibinfo {author}
  {\bibfnamefont {D.-H.}\ \bibnamefont {Huang}}, \bibinfo {author}
  {\bibfnamefont {J.}~\bibnamefont {Jiang}}, \bibinfo {author} {\bibfnamefont
  {P.}~\bibnamefont {Lalanne}},\ and\ \bibinfo {author} {\bibfnamefont
  {J.}~\bibnamefont {Fan}},\ }\bibfield  {title} {\bibinfo {title}
  {{Physics-augmented deep learning for high-speed electromagnetic simulation
  and optimization}},\ }\href
  {https://doi.org/10.21203/rs.3.rs-807786/v1%0ALicense:} {\bibfield  {journal}
  {\bibinfo  {journal} {PREPRINT available at Research Square}\ ,\ \bibinfo
  {pages} {1}} (\bibinfo {year} {2021})}\BibitemShut {NoStop}%
\bibitem [{\citenamefont {Lu}\ \emph {et~al.}(2021)\citenamefont {Lu},
  \citenamefont {Meng}, \citenamefont {Mao},\ and\ \citenamefont
  {Karniadakis}}]{Lu2021}%
  \BibitemOpen
  \bibfield  {author} {\bibinfo {author} {\bibfnamefont {L.}~\bibnamefont
  {Lu}}, \bibinfo {author} {\bibfnamefont {X.}~\bibnamefont {Meng}}, \bibinfo
  {author} {\bibfnamefont {Z.}~\bibnamefont {Mao}},\ and\ \bibinfo {author}
  {\bibfnamefont {G.~E.}\ \bibnamefont {Karniadakis}},\ }\bibfield  {title}
  {\bibinfo {title} {{DeepXDE: A deep learning library for solving differential
  equations}},\ }\href {https://doi.org/10.1137/19M1274067} {\bibfield
  {journal} {\bibinfo  {journal} {SIAM Review}\ }\textbf {\bibinfo {volume}
  {63}},\ \bibinfo {pages} {208} (\bibinfo {year} {2021})}\BibitemShut
  {NoStop}%
\bibitem [{\citenamefont {Chen}\ \emph {et~al.}(2020)\citenamefont {Chen},
  \citenamefont {Lu}, \citenamefont {Karniadakis},\ and\ \citenamefont {{Dal
  Negro}}}]{Chen2020c}%
  \BibitemOpen
  \bibfield  {author} {\bibinfo {author} {\bibfnamefont {Y.}~\bibnamefont
  {Chen}}, \bibinfo {author} {\bibfnamefont {L.}~\bibnamefont {Lu}}, \bibinfo
  {author} {\bibfnamefont {G.~E.}\ \bibnamefont {Karniadakis}},\ and\ \bibinfo
  {author} {\bibfnamefont {L.}~\bibnamefont {{Dal Negro}}},\ }\bibfield
  {title} {\bibinfo {title} {{Physics-informed neural networks for inverse
  problems in nano-optics and metamaterials}},\ }\href
  {https://doi.org/10.1364/oe.384875} {\bibfield  {journal} {\bibinfo
  {journal} {Optics Express}\ }\textbf {\bibinfo {volume} {28}},\ \bibinfo
  {pages} {11618} (\bibinfo {year} {2020})}\BibitemShut {NoStop}%
\bibitem [{\citenamefont {Chen}\ and\ \citenamefont {Negro}(2021)}]{Chen2021a}%
  \BibitemOpen
  \bibfield  {author} {\bibinfo {author} {\bibfnamefont {Y.}~\bibnamefont
  {Chen}}\ and\ \bibinfo {author} {\bibfnamefont {L.~D.}\ \bibnamefont
  {Negro}},\ }\bibfield  {title} {\bibinfo {title} {{Physics-informed neural
  networks for imaging and parameter retrieval of photonic nanostructures from
  near-field data}},\ }\href {https://doi.org/10.1063/5.0072969} {\bibfield
  {journal} {\bibinfo  {journal} {APL Photonics}\ }\textbf {\bibinfo {volume}
  {7}},\ \bibinfo {pages} {1} (\bibinfo {year} {2021})}\BibitemShut {NoStop}%
\bibitem [{\citenamefont {Lagaris}\ \emph {et~al.}(1998)\citenamefont
  {Lagaris}, \citenamefont {Likas},\ and\ \citenamefont
  {Fotiadis}}]{Lagaris1998}%
  \BibitemOpen
  \bibfield  {author} {\bibinfo {author} {\bibfnamefont {I.~E.}\ \bibnamefont
  {Lagaris}}, \bibinfo {author} {\bibfnamefont {A.}~\bibnamefont {Likas}},\
  and\ \bibinfo {author} {\bibfnamefont {D.~I.}\ \bibnamefont {Fotiadis}},\
  }\bibfield  {title} {\bibinfo {title} {{Artificial neural networks for
  solving ordinary and partial differential equations}},\ }\href
  {https://doi.org/10.1109/72.712178} {\bibfield  {journal} {\bibinfo
  {journal} {IEEE Transactions on Neural Networks}\ }\textbf {\bibinfo {volume}
  {9}},\ \bibinfo {pages} {987} (\bibinfo {year} {1998})}\BibitemShut {NoStop}%
\bibitem [{\citenamefont {Lim}\ and\ \citenamefont {Psaltis}(2022)}]{Lim2021}%
  \BibitemOpen
  \bibfield  {author} {\bibinfo {author} {\bibfnamefont {J.}~\bibnamefont
  {Lim}}\ and\ \bibinfo {author} {\bibfnamefont {D.}~\bibnamefont {Psaltis}},\
  }\bibfield  {title} {\bibinfo {title} {{MaxwellNet: Physics-driven deep
  neural network training based on Maxwell's equations}},\ }\href@noop {}
  {\bibfield  {journal} {\bibinfo  {journal} {APL Photonics}\ }\textbf
  {\bibinfo {volume} {7}},\ \bibinfo {pages} {1} (\bibinfo {year}
  {2022})}\BibitemShut {NoStop}%
\bibitem [{\citenamefont {Ronneberger}\ \emph {et~al.}(2015)\citenamefont
  {Ronneberger}, \citenamefont {Fischer},\ and\ \citenamefont
  {Brox}}]{Ronneberger}%
  \BibitemOpen
  \bibfield  {author} {\bibinfo {author} {\bibfnamefont {O.}~\bibnamefont
  {Ronneberger}}, \bibinfo {author} {\bibfnamefont {P.}~\bibnamefont
  {Fischer}},\ and\ \bibinfo {author} {\bibfnamefont {T.}~\bibnamefont
  {Brox}},\ }\bibfield  {title} {\bibinfo {title} {{U-Net: Convolutional
  Networks for Biomedical Image Segmentation}},\ }in\ \href@noop {} {\emph
  {\bibinfo {booktitle} {Medical Image Computing and Computer-Assisted
  Intervention -- MICCAI 2015}}},\ \bibinfo {editor} {edited by\ \bibinfo
  {editor} {\bibfnamefont {N.}~\bibnamefont {Navab}}, , \bibinfo {editor}
  {\bibfnamefont {J.}~\bibnamefont {Hornegger}}, , \bibinfo {editor}
  {\bibfnamefont {W.~M.}\ \bibnamefont {Wells}}, ,\ and\ \bibinfo {editor}
  {\bibfnamefont {A.~F.}\ \bibnamefont {Frangi}}}\ (\bibinfo  {publisher}
  {Springer International Publishing},\ \bibinfo {year} {2015})\ pp.\ \bibinfo
  {pages} {234--241}\BibitemShut {NoStop}%
\bibitem [{\citenamefont {Snyder}\ and\ \citenamefont
  {Love}(1983)}]{Snyder1983}%
  \BibitemOpen
  \bibfield  {author} {\bibinfo {author} {\bibfnamefont {A.}~\bibnamefont
  {Snyder}}\ and\ \bibinfo {author} {\bibfnamefont {J.}~\bibnamefont {Love}},\
  }\href {https://doi.org/10.1109/TCS.1979.1084602} {\emph {\bibinfo {title}
  {Chapman and Hall}}}\ (\bibinfo  {publisher} {Chapman and Hall},\ \bibinfo
  {year} {1983})\BibitemShut {NoStop}%
\bibitem [{\citenamefont {Yee}(1966)}]{Yee1966}%
  \BibitemOpen
  \bibfield  {author} {\bibinfo {author} {\bibfnamefont {K.~S.}\ \bibnamefont
  {Yee}},\ }\bibfield  {title} {\bibinfo {title} {{Numerical Solution of
  Initial Boundary Value Problems Involving Maxwell's Equations in Isotropic
  Media}},\ }\href {https://doi.org/10.1109/TAP.1966.1138693} {\bibfield
  {journal} {\bibinfo  {journal} {IEEE Transactions on Antennas and
  Propagation}\ }\textbf {\bibinfo {volume} {14}},\ \bibinfo {pages} {302}
  (\bibinfo {year} {1966})}\BibitemShut {NoStop}%
\bibitem [{\citenamefont {Chew}\ and\ \citenamefont {Weedon}(1994)}]{Chew1994}%
  \BibitemOpen
  \bibfield  {author} {\bibinfo {author} {\bibfnamefont {W.~C.}\ \bibnamefont
  {Chew}}\ and\ \bibinfo {author} {\bibfnamefont {W.~H.}\ \bibnamefont
  {Weedon}},\ }\bibfield  {title} {\bibinfo {title} {{A 3D perfectly matched
  medium from modified maxwell's equations with stretched coordinates}},\
  }\href {https://doi.org/10.1002/mop.4650071304} {\bibfield  {journal}
  {\bibinfo  {journal} {Microwave and Optical Technology Letters}\ }\textbf
  {\bibinfo {volume} {7}},\ \bibinfo {pages} {599} (\bibinfo {year}
  {1994})}\BibitemShut {NoStop}%
\bibitem [{\citenamefont {Johnson}(2021)}]{Johnson2021}%
  \BibitemOpen
  \bibfield  {author} {\bibinfo {author} {\bibfnamefont {S.~G.}\ \bibnamefont
  {Johnson}},\ }\bibfield  {title} {\bibinfo {title} {{Notes on Perfectly
  Matched Layers (PMLs)}},\ }\href {http://arxiv.org/abs/2108.05348} {\bibfield
   {journal} {\bibinfo  {journal} {arXiv:2108.0534}\ ,\ \bibinfo {pages} {1}}
  (\bibinfo {year} {2021})}\BibitemShut {NoStop}%
\bibitem [{\citenamefont {Farjadpour}\ \emph {et~al.}(2006)\citenamefont
  {Farjadpour}, \citenamefont {Roundy}, \citenamefont {Rodriguez},
  \citenamefont {Ibanescu}, \citenamefont {Bermel}, \citenamefont
  {Joannopoulos}, \citenamefont {Johnson},\ and\ \citenamefont
  {Burr}}]{Farjadpour2006}%
  \BibitemOpen
  \bibfield  {author} {\bibinfo {author} {\bibfnamefont {A.}~\bibnamefont
  {Farjadpour}}, \bibinfo {author} {\bibfnamefont {D.}~\bibnamefont {Roundy}},
  \bibinfo {author} {\bibfnamefont {A.}~\bibnamefont {Rodriguez}}, \bibinfo
  {author} {\bibfnamefont {M.}~\bibnamefont {Ibanescu}}, \bibinfo {author}
  {\bibfnamefont {P.}~\bibnamefont {Bermel}}, \bibinfo {author} {\bibfnamefont
  {J.~D.}\ \bibnamefont {Joannopoulos}}, \bibinfo {author} {\bibfnamefont
  {S.~G.}\ \bibnamefont {Johnson}},\ and\ \bibinfo {author} {\bibfnamefont
  {G.~W.}\ \bibnamefont {Burr}},\ }\bibfield  {title} {\bibinfo {title}
  {{Improving accuracy by subpixel smoothing in the finite-difference time
  domain}},\ }\href {https://doi.org/10.1364/ol.31.002972} {\bibfield
  {journal} {\bibinfo  {journal} {Optics Letters}\ }\textbf {\bibinfo {volume}
  {31}},\ \bibinfo {pages} {2972} (\bibinfo {year} {2006})}\BibitemShut
  {NoStop}%
\bibitem [{\citenamefont {Kang}\ \emph {et~al.}(2021)\citenamefont {Kang},
  \citenamefont {Uchida},\ and\ \citenamefont {Iwana}}]{Kang2021}%
  \BibitemOpen
  \bibfield  {author} {\bibinfo {author} {\bibfnamefont {S.}~\bibnamefont
  {Kang}}, \bibinfo {author} {\bibfnamefont {S.}~\bibnamefont {Uchida}},\ and\
  \bibinfo {author} {\bibfnamefont {B.~K.}\ \bibnamefont {Iwana}},\ }\bibfield
  {title} {\bibinfo {title} {{Tunable U-Net: Controlling Image-to-Image Outputs
  Using a Tunable Scalar Value}},\ }\href
  {https://doi.org/10.1109/ACCESS.2021.3096530} {\bibfield  {journal} {\bibinfo
   {journal} {IEEE Access}\ }\textbf {\bibinfo {volume} {9}},\ \bibinfo {pages}
  {103279} (\bibinfo {year} {2021})}\BibitemShut {NoStop}%
\bibitem [{\citenamefont {Weber}(2003)}]{Weber2003}%
  \BibitemOpen
  \bibfield  {author} {\bibinfo {author} {\bibfnamefont {M.~J.}\ \bibnamefont
  {Weber}},\ }\href {https://doi.org/10.5860/choice.40-4668} {\emph {\bibinfo
  {title} {Handbook of optical materials}}},\ edited by\ \bibinfo {editor}
  {\bibfnamefont {M.~J.}\ \bibnamefont {Weber}}\ (\bibinfo  {publisher} {CRC
  PRESS},\ \bibinfo {year} {2003})\BibitemShut {NoStop}%
\bibitem [{\citenamefont {Adair}\ \emph {et~al.}(1987)\citenamefont {Adair},
  \citenamefont {Chase},\ and\ \citenamefont {Payne}}]{Adair1987}%
  \BibitemOpen
  \bibfield  {author} {\bibinfo {author} {\bibfnamefont {R.}~\bibnamefont
  {Adair}}, \bibinfo {author} {\bibfnamefont {L.~L.}\ \bibnamefont {Chase}},\
  and\ \bibinfo {author} {\bibfnamefont {S.~A.}\ \bibnamefont {Payne}},\
  }\bibfield  {title} {\bibinfo {title} {{Nonlinear Refractive-Index
  Measurements of Glasses and Crystals Using Three-Wave Frequency Mixing.}},\
  }\href@noop {} {\bibfield  {journal} {\bibinfo  {journal} {JOSA B}\ }\textbf
  {\bibinfo {volume} {4}},\ \bibinfo {pages} {875} (\bibinfo {year}
  {1987})}\BibitemShut {NoStop}%
\bibitem [{\citenamefont {Jiang}\ and\ \citenamefont {Fan}(2019)}]{Jiang2019c}%
  \BibitemOpen
  \bibfield  {author} {\bibinfo {author} {\bibfnamefont {J.}~\bibnamefont
  {Jiang}}\ and\ \bibinfo {author} {\bibfnamefont {J.~A.}\ \bibnamefont
  {Fan}},\ }\bibfield  {title} {\bibinfo {title} {{Global Optimization of
  Dielectric Metasurfaces Using a Physics-Driven Neural Network}},\ }\href
  {https://doi.org/10.1021/acs.nanolett.9b01857} {\bibfield  {journal}
  {\bibinfo  {journal} {Nano Letters}\ }\textbf {\bibinfo {volume} {19}},\
  \bibinfo {pages} {5366} (\bibinfo {year} {2019})}\BibitemShut {NoStop}%
\bibitem [{\citenamefont {Chen}\ and\ \citenamefont {Ahmed}(2021)}]{Chen2021b}%
  \BibitemOpen
  \bibfield  {author} {\bibinfo {author} {\bibfnamefont {W.}~\bibnamefont
  {Chen}}\ and\ \bibinfo {author} {\bibfnamefont {F.}~\bibnamefont {Ahmed}},\
  }\bibfield  {title} {\bibinfo {title} {{MO-PaDGAN: Reparameterizing
  Engineering Designs for augmented multi-objective optimization}},\ }\href
  {https://doi.org/10.1016/j.asoc.2021.107909} {\bibfield  {journal} {\bibinfo
  {journal} {Applied Soft Computing}\ }\textbf {\bibinfo {volume} {113}},\
  \bibinfo {pages} {1} (\bibinfo {year} {2021})}\BibitemShut {NoStop}%
\end{thebibliography}%

\end{document}